\documentclass[10pt,conference]{IEEEtran}
\usepackage[left=1.62cm,right=1.62cm,top=1.9cm]{geometry}

\IEEEoverridecommandlockouts
\usepackage{amsmath,amssymb,amsfonts}
\usepackage{algorithm}
\usepackage{algpseudocode}
\usepackage{graphicx}
\usepackage{textcomp}
\usepackage{xcolor}
\usepackage{balance}       
\usepackage{graphics}      
\usepackage[T1]{fontenc}   
\usepackage{txfonts}
\usepackage{mathptmx}
\usepackage[pdflang={en-US},pdftex]{hyperref}
\usepackage{color}
\usepackage{booktabs}
\usepackage{textcomp}
\usepackage{hyperref}       
\usepackage{url}            
\usepackage{multirow}
\usepackage{subcaption}

\usepackage{microtype}        
\usepackage{ccicons}          

\usepackage{todonotes}

\begin{document}

\title{Real-time Food Recipe Generation with \\Mobile Deep Learning and Large Language Model}

\author{
\IEEEauthorblockN{Wei Jun Leung, Mun Hong Lee, Joo Wee Quek, Desmond Loy, Yifei Wang, Yan Ning Ang, \\Pai Chet Ng, Kan Chen, Jeannie Lee}
\IEEEauthorblockA{
Infocomm Technology Cluster, Singapore Institute of Technology, Singapore \\
\{2201316, 2200634,2201046,2201435,2201777,2203428\}@sit.singaporetech.edu.sg, \\ \{paichet.ng, kan.chen, jeannie.lee\}@singaporetech.edu.sg}
}

\maketitle

\begin{abstract}
Finding recipes that match available kitchen ingredients can be cumbersome and inefficient. This paper addresses this challenge by developing a mobile application that generates tailored food recipes based on ingredients identified using a mobile camera. Leveraging deep learning and large language model (LLM), our system offers a seamless and efficient solution for real-time recipe generation. By integrating deep learning for accurate ingredient recognition with LLM for generating detailed and practical recipes, our approach ensures users receive precise cooking instructions based on actual ingredients on hand. Extensive dataset augmentation and model training have resulted in a robust system capable of high performance. Experimental results demonstrate that EfficientNet, used for food recognition, offers a balance of accuracy and computational efficiency, making it ideal for mobile applications. This dual-approach method enhances the cooking experience, reduces food waste, and provides a comprehensive, real-time tool for everyday cooking needs, bridging a significant gap in current food recommendation systems.
\end{abstract}


\section{Introduction}
In today's fast-paced world, mobile devices have become an integral part of our daily lives, providing convenience and accessibility to a plethora of services and information. One common use case for mobile devices is searching for food recipes online. However, this process can often be cumbersome and inefficient. Many online recipes require ingredients that may not be readily available in our kitchens, necessitating a trip to the grocery store and adding an extra layer of complexity to meal preparation. This inconvenience is compounded when users face the challenge of navigating through a vast array of recipe websites, each with different media formats and varying levels of reliability. Moreover, the time spent searching for recipes that match the ingredients on hand can be frustrating, especially for individuals with busy schedules or limited culinary experience. Recent advancements in machine learning and artificial intelligence have highlighted the potential for more intuitive and personalized solutions, yet many current applications still fall short in addressing the practical needs of everyday users. Integrating seamless, intelligent recipe generation into a multimedia system for mobile devices represents a significant step forward in making cooking more accessible and enjoyable for everyone.

The motivation behind our paper is to simplify this process by developing a mobile application that generates food recipes based on the ingredients available in the user's kitchen. This eliminates the need for extensive searches and makes cooking more accessible and convenient. While previous works \cite{Papadopoulos_2022_CVPR, 10.1145.2964284.2964315, 10.1145.3542954.3542983} have made strides in ingredient recognition and recipe retrieval using deep learning models, our approach uniquely integrates deep learning for ingredient identification with large language model (LLM) for detailed recipe generation. This dual approach ensures that users receive accurate and practical cooking instructions based on the actual ingredients they have on hand, filling a critical gap in the existing literature on food recommendation systems.

By leveraging the ubiquitous nature of mobile devices, we aim to provide a seamless experience where users can simply take a photo of their ingredients using their mobile camera and receive tailored recipe suggestions. Our objective is to create an intelligent system that combines food recognition with deep learning and recipe generation with LLM. The deep learning model accurately identifies the ingredients from the captured image, while the LLM generates relevant recipes based on these recognized ingredients. To ensure the accuracy and relevance of the recipes, we provide the LLM with detailed information about the available ingredients and nutritional expectations, guiding it to produce practical and accurate cooking instructions. This dual approach is novel because it integrates real-time ingredient recognition with advanced recipe generation, a combination that has not been fully addressed by existing solutions. This innovative method addresses the practical challenge of utilizing available ingredients effectively, making cooking more accessible and reducing food waste, thus providing a significant improvement over current recipe search methods.
We highlight our contributions as follows:
\begin{itemize}
    \item \textbf{Dataset Augmentation:} We augmented the GroceryStoreDataset by merging fine-grained categories into broader classes, such as grouping different apple varieties into a single “Apple” class. We further expanded the dataset with images scraped from Google and samples from Fruits-360, resulting in approximately 42,000 images split into training, validation, and test sets at a 70:15:15 ratio.
    \item \textbf{Integration of Deep Learning and LLM:} Our system integrates deep learning-based food recognition with an LLM for recipe generation. The deep learning model identifies ingredients from captured images, and the LLM uses them to generate relevant, practical recipes, providing users with cooking instructions based on the ingredients they actually have.
    \item \textbf{Extensive Experimentation and Validation:} We conducted experiments with various pre-trained deep learning and language models to evaluate the integrated system, including the accuracy and efficiency of food recognition and the relevance and practicality of LLM-generated recipes. The results demonstrate the robustness, effectiveness, and practical potential of our approach.
    \item \textbf{Full-Stack Mobile Application with User-Friendly Interface:} We developed a full-stack mobile application with a user-friendly interface. The Android Studio front end supports image capture, recipe browsing, favorites, and profile management, while the Flask- and Docker-based back end handles REST APIs, image classification, external API calls, and MongoDB operations to ensure a seamless user experience.
\end{itemize}

\section{Related Work}
\noindent
\textbf{Food Recognition:}
Morol et al. implemented a model for food ingredient recognition and designed an algorithm for recommending recipes based on recognized ingredients using a custom dataset of 9856 images \cite{10.1145/3542954.3542983}. Ming et al.  proposed a system applying image recognition to recognize food and analyze nutrition for health management \cite{ming2018food}. Chen and Ngo proposed deep architectures for simultaneous learning of ingredient recognition and food categorization, addressing the challenges of large visual variations in food appearance and ingredient composition \cite{10.1145/2964284.2964315}. Further advancements were made by Chen et al.  through multi-task and region-wise deep learning for ingredient recognition, focusing on the complexity of food ingredient identification at both image-level and region-level scales \cite{9305995}. Gao et al. proposed combining a food-orient Transformer and multi-task category-ingredient recognition network, focusing on ingredient relationships \cite{gao2019hierarchical}. However, these approaches primarily focus on ingredient detection and categorization without integrating comprehensive recipe generation capabilities.

\noindent
\textbf{Recipe Generation:}
Papadopoulos et al. introduced a model that learns a joint embedding between recipes and food images, projecting image-recipe embeddings into structured programs \cite{Papadopoulos_2022_CVPR}. Wang et al. proposed methods to learn structural representations for lengthy recipes to enhance recipe generation and food cross-modal retrieval \cite{9793670}. Similarly, Zhu et al. explored cross-modal recipe retrieval by leveraging paired recipe-image data to train deep models for food recognition and recipe retrieval \cite{9594708}. These methods are effective, however they often depend heavily on large and diverse datasets and do not fully leverage the potential of modern LLM for generating detailed and context-aware recipes.

\noindent
\textbf{LLMs for Food-related Application:}
Pitsilou et al. explored using LLMs to extract food entities from cooking recipes, emphasizing prompt and response engineering to minimize the need for labeled instances \cite{10555071}. Rodríguez-de-Vera et al. introduced the Dining on Details framework, which utilizes LLMs for fine-grained food recognition by constructing subsets of classes within datasets \cite{10.1145/3607828.3617797}. Zhou et al. developed FoodSky, a food-oriented LLM that excels in both chef and dietetic examinations, demonstrating its capability to understand and generate food-related content \cite{zhou2024foodskyfoodorientedlargelanguage}. More recently, Liu et al. proposed a profile-aware recipe generation framework that employs a Planner-Critic architecture to generate recipes from food images while incorporating user-specific dietary restrictions and safety constraints. Their approach introduces an iterative remediation mechanism to identify and revise unsafe recipes, highlighting the importance of trustworthy and personalized recipe generation \cite{liu2026profile}. While these studies demonstrate the potential of LLMs for food understanding and recipe generation, our work focuses on integrating deep learning-based ingredient recognition with LLM-driven recipe generation in a practical mobile application, enabling users to generate recipes directly from available kitchen ingredients.

\section{Proposed Dual-approach Method}
Our proposed solution leverages a dual-approach method that integrates deep learning for food recognition and LLM for recipe generation. The goal is to provide users with tailored recipe suggestions based on the ingredients they have on hand, using a seamless and intuitive mobile application interface.

\begin{algorithm}[t!]
\caption{Dual-Approach Method for Recipe Generation}
\begin{algorithmic}[1]
\State \textbf{Input:} Image $x$
\State \textbf{Output:} Recipe $R$
\Procedure{GenerateRecipe}{$x$}
    \State Capture image of ingredients ($x$) using mobile camera.
    \State Perform ingredient recognition to identify ingredients \[ y = \{ y_1, y_2, \ldots, y_n \} \]
    \State Retrieve top-$k$ nutrition from nutrition database \[ Z = \{ z_1, z_2, \ldots, z_k \} \]
    \State Craft prompt $P$ based on  $y$ and selected recipe $z_i$
    \State Use LLM to generate recipe $R$ based on prompt $P$:
        \[
        R = \{ \text{Step}_1, \text{Step}_2, \ldots, \text{Step}_m \}
        \]
    \State \Return $R$
\EndProcedure
\end{algorithmic}
\end{algorithm}

The dual-approach method described in Algorithm 1 integrates deep learning and LLM to provide an intelligent system for food recipe generation. For our proposed system solution, users begin by capturing an image of the ingredients available in their kitchen using their mobile camera, denoted as $ x$. An image recognition model is then used to identify the ingredients present in the image. In this work, we adapt the \textit{EfficientNet} \cite{tan2019efficientnet} to perform this food recognition task due to its high accuracy and efficiency. The output of this model is a set of recognized ingredients $ y$, where $ y = \{ y_1, y_2, \ldots, y_n \}$. Each $ y_i$ represents a distinct ingredient identified in the image.

Once the ingredients $y$ are identified, the system invokes an API call to a nutrition database, here we used the database from Edamam \cite{edamam_api}, to retrieve a list of potential nutrition that can be prepared using these ingredients. Let $Z = \{ z_1, z_2, \ldots, z_k \}$ be the set of top-$k$ recommended food from the retrieved nutrition. Each $z_i$ represents a recommended food with potential nutrition that matches the available ingredients $y$.

To generate accurate and practical cooking instructions, the system crafts a detailed prompt for the LLM based on the recognized ingredients and selected recipe. The prompt $P$ is formulated to include the list of recognized ingredients $y$, nutritional expectations, and dietary requirements, along with the selected food from the recommendation set $Z$. The prompt $P$ can be expressed as: 

\begin{quote}
\color{gray!80}
Generate a recipe for the selected food, $z_i$, using the following ingredients: $y_1, y_2, \ldots, y_n$. Consider the nutritional values and dietary preferences.
\end{quote}

The LLM, here we consider OpenAI’s GPT-3.5 \cite{openai_gpt35}, processes the prompt $P$ to generate a realistic and detailed recipe $R$. The output $R$ includes the step-by-step instructions for preparing the dish, ensuring that the recipe is practical and based on the ingredients provided. The generated recipe $R$ can be represented as: $R = \{ \text{Step}_1, \text{Step}_2, \ldots, \text{Step}_m \}$.

By combining deep learning for ingredient recognition with an LLM for recipe generation, our system provides a novel and comprehensive solution that has not been fully addressed by existing methods. This architecture ensures that users receive accurate and practical recipes based on the ingredients they have, enhancing the cooking experience and reducing the need for additional grocery shopping. This dual-approach method is encapsulated in a scalable and containerized backend infrastructure, ensuring robust performance and user convenience. The architecture not only demonstrates the feasibility of integrating advanced AI technologies but also addresses real-world challenges in food preparation and recipe generation.


\begin{figure*}[t!]
  \centering
  \includegraphics[width=.8\linewidth]{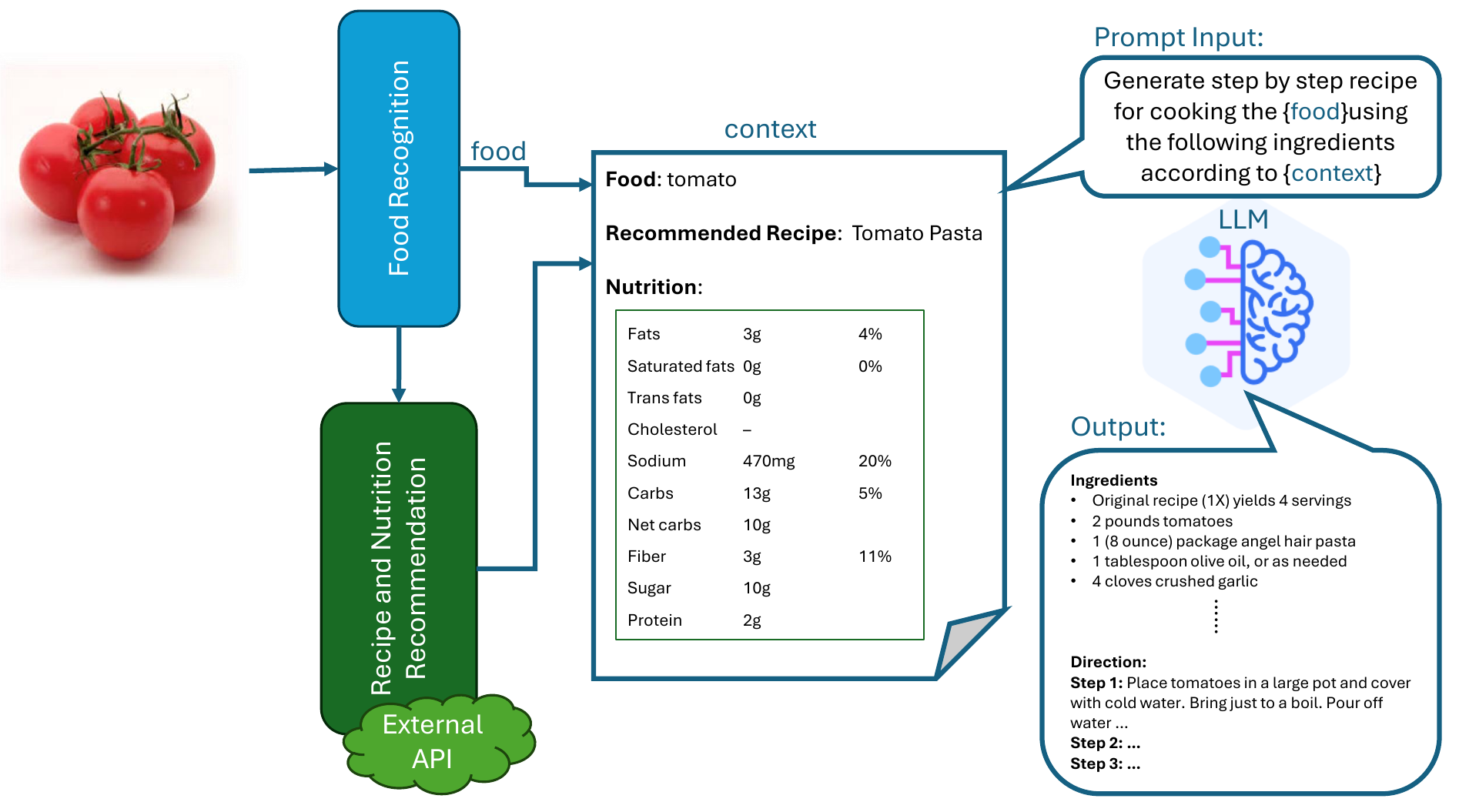}
  \caption{Deep learning was employed for food recognition, while an external API facilitated ingredient recommendations. An LLM was integrated to generate recipes based on the identified food, incorporating suggested ingredients and nutritional guidelines.}
  \label{fig:architecture}
\end{figure*}
\section{System Implementation}
As shown in Fig.~\ref{fig:architecture}, Our system implementation involves the development of a full-stack mobile application designed to recognize ingredients from images, recommend suitable food recipe, and generate detailed cooking instructions. 
\begin{figure}[t!]
  \centering
  \includegraphics[width=\linewidth]{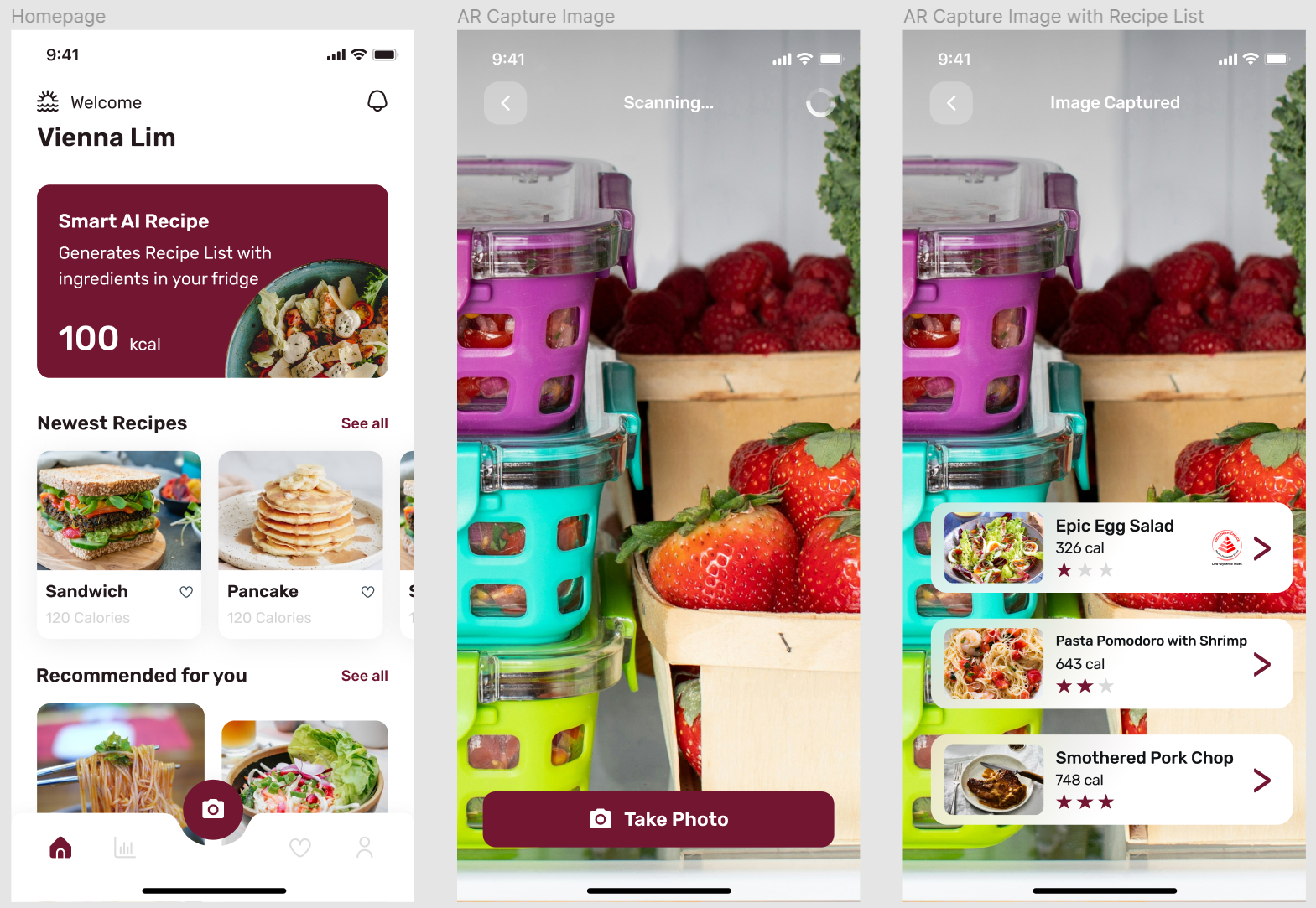}
  \caption{Home, Camera and Recommendation pages.}
  \label{fig:recommendation}
\end{figure}

\begin{figure}[t!]
  \centering
  \includegraphics[width=\linewidth]{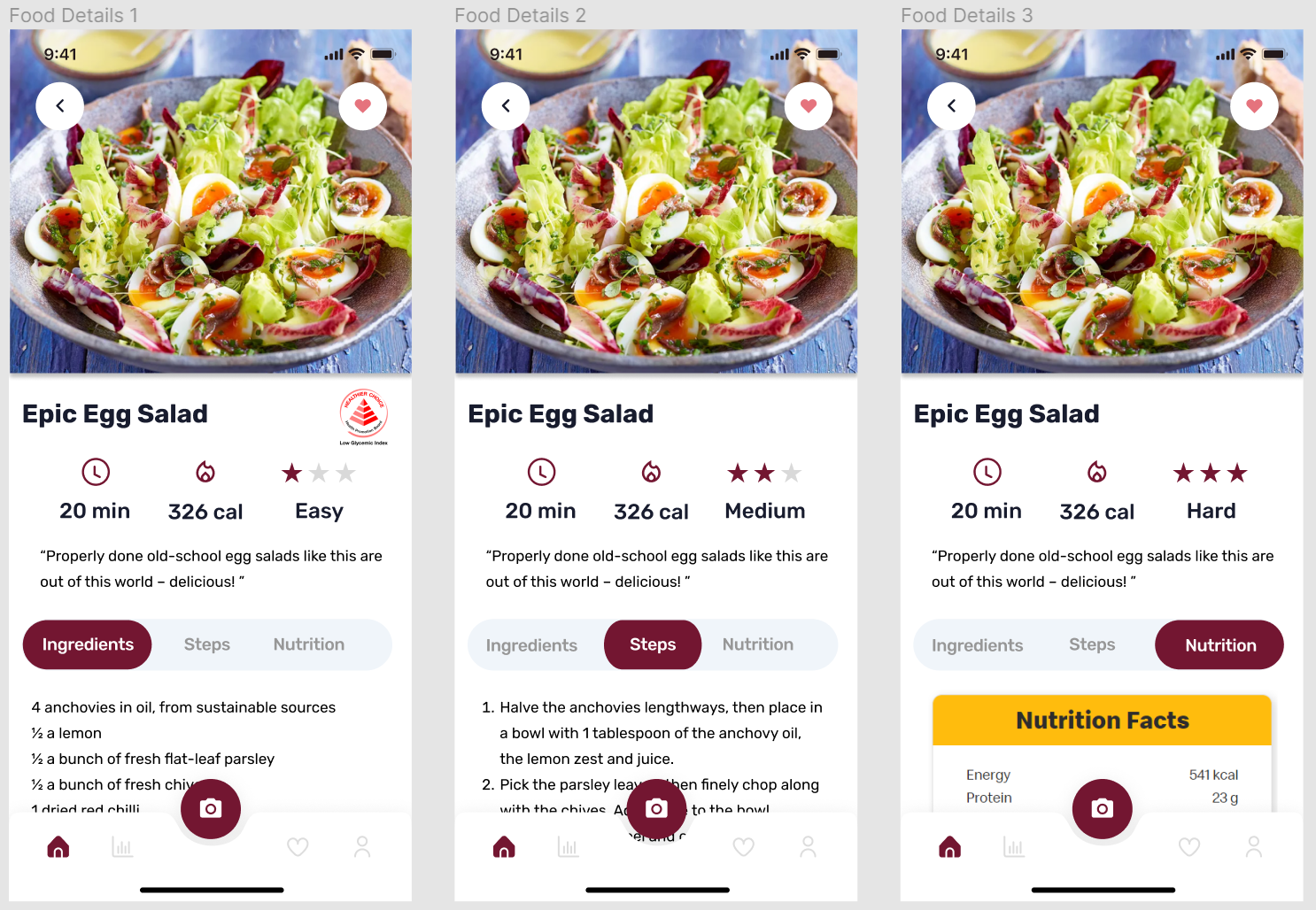}
  \caption{Recipe Information page}
  \label{fig:recipe}
\end{figure}
\subsection{Front-End Mobile Application}
The front end of our mobile application is developed using Android Studio, providing a user-friendly interface for capturing images of ingredients and interacting with recipe suggestions. Upon opening the application, users are greeted with an onboarding screen, where they can either sign in to an existing registered account or sign up for a new one. After logging in, users are directed to the homepage, which displays the newest recipes as well as personalized recommendations. Users can capture an image of the ingredients available in their kitchen by clicking the camera icon in the bottom navigation bar. After an image is captured, the application utilizes AI to suggest suitable food options. As shown in Fig.~\ref{fig:recommendation}, the recommended foods are displayed in a list on the screen, allowing users to select their desired food. Once a food item is selected, as shown in Fig.~\ref{fig:recipe}, users can view detailed information, including the required ingredients, step-by-step cooking instructions, and nutritional facts. The application also allows users to save their favorite recipes for future reference and update their personal information through the profile page. In addition, the history page enables users to revisit recipes they have previously viewed.

\subsection{Back-End Server}
The back-end server, implemented using Flask, handles the core functionalities of the application, including image classification (specifically food recognition), food recommendation, and interaction with external APIs to retrieve the nutrition and generate the cooking instructions. 

\subsubsection{Food Recognition Model}
We adapt an image classification model and fine-tune it to work as food recognition.
The food recognition model forms the backbone of our application, enabling the recognition of ingredients from captured images. We selected \textit{EfficientNet} \cite{tan2019efficientnet} for its balance of speed and accuracy. We have also compared our model performance with a few baselines models including AlexNet \cite{krizhevsky2012imagenet}, VGG16 \cite{simonyan2014very}, and ResNet50 \cite{he2016deep}, and the performance comparison is presented in the next section.

\subsubsection{External API Integration}
The integration of external APIs is a critical aspect of our system, enabling comprehensive functionality. The Edamam API is used for recipe recommendation, providing a vast database of recipes and nutritional information. Additionally, the system utilizes OpenAI’s GPT-3.5 turbo model for generating detailed cooking instructions. The crafted prompt, which includes the recognized ingredients and selected recipe, is fed to the LLM, resulting in a realistic and practical recipe output.

\paragraph{Food Recipe Recommendation}
Upon recognizing the ingredients, the back-end server invokes the Edamam API \cite{edamam_api} to retrieve a list of potential food recipes and the corresponding nutrition. The Edamam API offers a comprehensive database of over 2.3 million recipes, including information on calories and nutritional values. It also allows the filtering of recipes based on dietary requirements. The server uses the recognized ingredients to query the Edamam API, obtaining a set of top-$k$ recommended recipes.

\paragraph{Cooking instructions Generation with LLM}
To generate accurate and practical cooking instructions, the back-end server formulates a detailed prompt for the LLM (in our case, we used OpenAI's GPT-3.5 turbo \cite{openai_gpt35}) based on the recognized ingredients and selected recipe. The prompt includes the list of ingredients, nutritional expectations, and dietary preferences. The formulated prompt ensures that the LLM generates a recipe that is both realistic and tailored to the user's needs.

\subsection{Database Management}
To facilitate the storage and retrieval of user data, including favorite recipes, we use MongoDB \cite{mongodb}. The database stores recipes and their preparation steps as documents, tagged to user IDs. This allows users to save and access their favorite recipes at any time, enhancing the overall user experience.
By combining deep learning for ingredient recognition with LLM for recipe generation, our system provides a comprehensive solution that addresses real-world challenges in food preparation and recipe generation. This architecture ensures that users receive accurate and practical recipes based on the ingredients they have, making cooking more accessible and reducing the need for additional grocery shopping.

\section{Experiment and Evaluation}

\subsection{Dataset Augmentation}

As a proof of concept, we chose the GroceryStoreDataset \cite{klasson2019hierarchical}. This dataset presented some challenges as it divided each fruit/item into finer classes (e.g., Apple: Golden-Delicious, Granny-Smith). To address this, we developed a Python script to merge these finer classes into generic ones (e.g., Apple). Building on this, we supplemented the dataset by scraping images from Google and incorporating the Fruits-360 dataset \cite{muresan2018fruit}. Ultimately, we compiled a dataset of approximately 42,000 images with a train:validation:test split ratio of 70:15:15.
Our model currently has 75 class labels, encompassing a variety of grocery items such as vegetables, fruits, meat, and some packaged goods (e.g., milk cartons). These class labels are largely inherited from the GroceryStoreDataset (approximately 42 classes), with additional labels based on the datasets we scraped from the web. 

\subsection{Baseline Models}

For evaluation and comparison, we selected three state-of-the-art image classification models: AlexNet \cite{krizhevsky2012imagenet}, VGG16 \cite{simonyan2014very}, and ResNet50 \cite{he2016deep}. These models were chosen for their well-established performance in image classification tasks. Each model was fine-tuned using the same augmented dataset to ensure a fair comparison. 

\subsection{Evaluation Metrics}
The evaluation of the models was based on three main criteria: model performance, model complexity, and inference time.
\begin{itemize}
    \item \textbf{Model Performance}: This was measured in terms of accuracy, precision, recall, and F1-score. These metrics provide a comprehensive view of each model's ability to correctly classify images.
    \item \textbf{Model Complexity}: This was evaluated based on the number of parameters and the MACs/FLOPs (multiply-accumulate operations/floating point operations per second). These metrics help in understanding the computational requirements of each model.
    \item \textbf{Inference Time}: The time taken by each model to perform inference on a given image was recorded to assess the practical usability of the models in a real-time application.
\end{itemize}

\begin{table}[t]
    \centering
    \caption{Model Performance}
    \setlength{\tabcolsep}{4pt}
    \resizebox{\linewidth}{!}{
    \begin{tabular}{|l|c|c|c|c|}
        \hline
        \textbf{Metric / Model} & \textbf{AlexNet} & \textbf{VGG16} & \textbf{ResNet50} & \textbf{EfficientNet} \\ \hline
        \textbf{Accuracy} & 87.49\% & 60.65\% & 98.73\% & 85.74\% \\ \hline
        \textbf{Precision} & 91.67\% & 66.43\% & 98.88\% & 90.20\% \\ \hline
        \textbf{Recall} & 87.49\% & 60.65\% & 98.73\% & 85.74\% \\ \hline
        \textbf{F1-Score} & 86.74\% & 61.40\% & 98.72\% & 85.65\% \\ \hline
    \end{tabular}}
    \label{tab:model_performance}
\end{table}

\begin{table}[t]
    \centering
    \caption{Model Complexity}
    \setlength{\tabcolsep}{4pt}
    \resizebox{\linewidth}{!}{
    \begin{tabular}{|l|c|c|c|c|}
        \hline
        \textbf{Metric / Model} & \textbf{AlexNet} & \textbf{VGG16} & \textbf{ResNet50} & \textbf{EfficientNet} \\ \hline
        \textbf{MACs/FLOPs} & 710,647,488 & 15,466,755,072 & 4,131,962,880 & 414,032,704 \\ \hline
        \textbf{Params} & 57,540,547 & 134,797,251 & 23,776,451 & 4,175,359 \\ \hline
    \end{tabular}}
    \label{tab:model_complexity}
\end{table}

\begin{figure}[t]
  \centering
  \includegraphics[width=\linewidth]{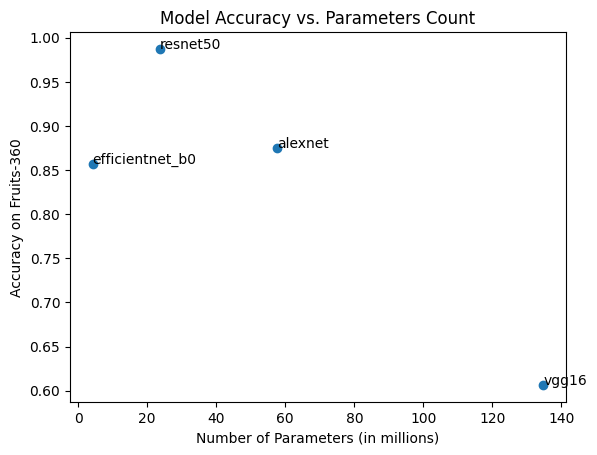}
  \caption{Parameter count vs Testing Accuracy}
  \label{fig:modelAccuracy}
\end{figure}

\subsection{Results and Discussion}

\subsubsection{Model Performance}
The model performance metrics are summarized in Table \ref{tab:model_performance}. ResNet50 achieved the highest accuracy, precision, recall, and F1-score, indicating its strong performance in classifying images. However, EfficientNet also demonstrated competitive performance with an accuracy of 85.74\%, precision of 90.20\%, recall of 85.74\%, and F1-score of 85.65\%. While it does not surpass ResNet50, EfficientNet offers a good balance of accuracy and other performance metrics, making it a viable option for our application.

\subsubsection{Model Complexity}
Model complexity is evaluated based on the number of parameters and MACs/FLOPs, as shown in Table \ref{tab:model_complexity}. ResNet50, despite its high performance, has a significantly higher number of parameters and MACs/FLOPs compared to EfficientNet. This is further illustrated in Fig. \ref{fig:modelAccuracy}, which compares the number of parameters with the testing accuracy. EfficientNet stands out with a much lower parameter count and MACs/FLOPs, while still maintaining a competitive accuracy. This balance of performance and complexity makes EfficientNet more suitable for our mobile application, which requires a lightweight model that can run efficiently on mobile devices.

\subsubsection{Inference Time}
\begin{figure}[t]
    \centering
    \begin{subfigure}[b]{0.45\linewidth}
        \centering
        \includegraphics[width=\linewidth]{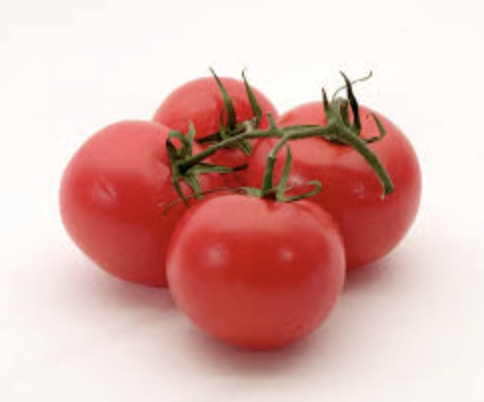}
        \caption{}
    \end{subfigure}
    \hspace{0.5cm}
    \begin{subfigure}[b]{0.45\linewidth}
        \centering
        \includegraphics[width=\linewidth]{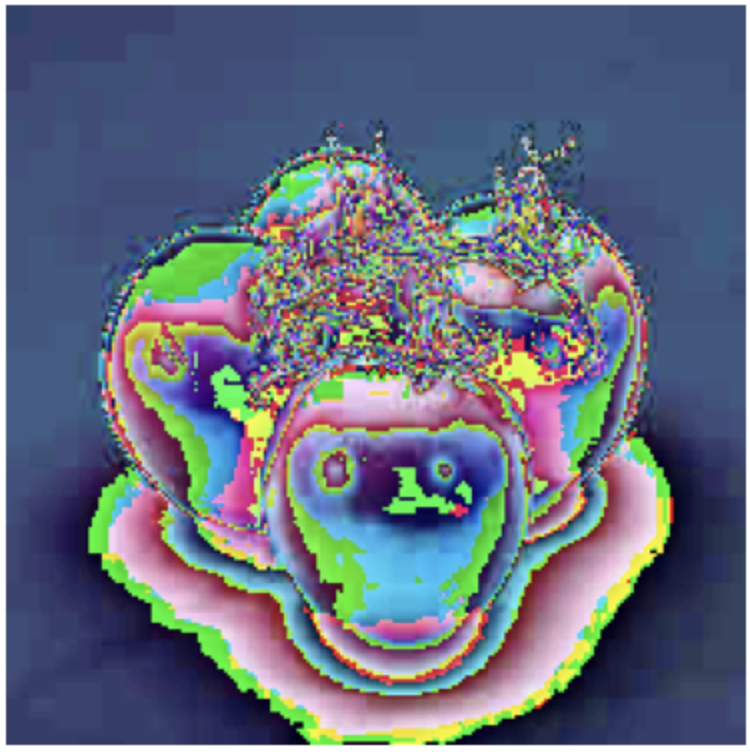}
        \caption{}
    \end{subfigure}
    \caption{Original (a) and processed (b) tomato images.}
    \label{fig:original_tomato}
\end{figure}

Inference time was evaluated using real images captured by our mobile app. For example, Fig. \ref{fig:original_tomato} (a) shows the original tomato image that was captured, while Fig. \ref{fig:original_tomato} (b) shows the processed image. Table \ref{tab:model_inference} summarizes the inference times and predicted class labels for each model. EfficientNet had an inference time of 0.010387 seconds, which is reasonable for real-time applications. Although ResNet50 had a higher test accuracy, its inference time was longer, and it incorrectly predicted the tomato image as "Pear Kaiser." This indicates that the model may be overfitted and not perform well on unseen images. This highlights the importance of balancing model complexity and performance, especially for real-time applications where inference speed is crucial.

\begin{table}[t]
    \centering
    \caption{Model Performance on Google Images}
    \resizebox{\linewidth}{!}{
    \begin{tabular}{|l|l|r|}
        \hline
        \textbf{Model} & \textbf{Predicted Label} & \textbf{Inference Time (seconds)} \\ \hline
        AlexNet & Tomato 2 & 0.002084s \\ \hline
        EfficientNet & Tomato 2 & 0.010387s \\ \hline
        VGG16 & Tomato 2 & 0.008836s \\ \hline
        ResNet50 & Pear Kaiser & 0.011327s \\ \hline
    \end{tabular}}
    \label{tab:model_inference}
\end{table}


\section{Conclusion}
This paper demonstrates the effectiveness of integrating deep learning and LLM for generating food recipes based on real-time ingredient recognition via mobile devices. EfficientNet was selected for its balance of performance and complexity, achieving competitive accuracy with significantly fewer parameters and lower computational requirements compared to other models like ResNet50. This model's inference time is suitable for real-time applications, and it performed well on unseen images, indicating robustness and practical usability. Future work could explore expanding the dataset to include a broader range of ingredients and improving the accuracy of ingredient recognition. Additionally, integrating user feedback to refine recipe suggestions and incorporating more advanced dietary and nutritional considerations can further enhance the system's utility. Advanced LLM could be employed to generate more complex and diverse recipes, making the application even more versatile and user-friendly.


\bibliographystyle{IEEEbib}
\bibliography{ref}

@String{Computing = "Computing" }

@String{Computer = "{IEEE} Computer" }

@String{Springer = "Springer-Verlag" }

@inproceedings{10.1145.3542954.3542983,
author = {Morol, Md. Kishor and Rokon, Md. Shafaat Jamil and Hasan, Ishra Binte and Saif, A. M. and Khan, Rafid Hussain and Das, Shuvra Smaran},
title = {Food Recipe Recommendation Based on Ingredients Detection Using Deep Learning},
year = {2022},
isbn = {9781450397346},
publisher = {Association for Computing Machinery},
address = {New York, NY, USA},
url = {https://doi.org/10.1145/3542954.3542983},
doi = {10.1145/3542954.3542983},
booktitle = {Proceedings of the 2nd International Conference on Computing Advancements},
pages = {191–198},
numpages = {8},
location = {Dhaka, Bangladesh},
series = {ICCA '22}
}

@inproceedings{10.1145.2964284.2964315,
author = {Chen, Jingjing and Ngo, Chong-wah},
title = {Deep-based Ingredient Recognition for Cooking Recipe Retrieval},
year = {2016},
isbn = {9781450336031},
publisher = {Association for Computing Machinery},
address = {New York, NY, USA},
url = {https://doi.org/10.1145/2964284.2964315},
doi = {10.1145/2964284.2964315},
booktitle = {Proceedings of the 24th ACM International Conference on Multimedia},
pages = {32–41},
numpages = {10},
location = {Amsterdam, The Netherlands},
series = {MM '16}
}

@ARTICLE{9305995,
  author={Chen, Jingjing and Zhu, Bin and Ngo, Chong-Wah and Chua, Tat-Seng and Jiang, Yu-Gang},
  journal={IEEE Transactions on Image Processing}, 
  title={A Study of Multi-Task and Region-Wise Deep Learning for Food Ingredient Recognition}, 
  year={2021},
  volume={30},
  number={},
  pages={1514-1526},
  doi={10.1109/TIP.2020.3045639},
  ISSN={1941-0042},
  month={},}

@InProceedings{Papadopoulos_2022_CVPR,
    author    = {Papadopoulos, Dim P. and Mora, Enrique and Chepurko, Nadiia and Huang, Kuan Wei and Ofli, Ferda and Torralba, Antonio},
    title     = {Learning Program Representations for Food Images and Cooking Recipes},
    booktitle = {Proceedings of the IEEE/CVF Conference on Computer Vision and Pattern Recognition (CVPR)},
    month     = {June},
    year      = {2022},
    pages     = {16559-16569}
}

@ARTICLE{9793670,
  author={Wang, Hao and Lin, Guosheng and Hoi, Steven C. H. and Miao, Chunyan},
  journal={IEEE Transactions on Pattern Analysis and Machine Intelligence}, 
  title={Learning Structural Representations for Recipe Generation and Food Retrieval}, 
  year={2023},
  volume={45},
  number={3},
  pages={3363-3377},
  doi={10.1109/TPAMI.2022.3181294},
  ISSN={1939-3539},
  month={March},}

@ARTICLE{9594708,
  author={Zhu, Bin and Ngo, Chong-Wah and Chan, Wing-Kwong},
  journal={IEEE Transactions on Multimedia}, 
  title={Learning From Web Recipe-Image Pairs for Food Recognition: Problem, Baselines and Performance}, 
  year={2022},
  volume={24},
  number={},
  pages={1175-1185},
  doi={10.1109/TMM.2021.3123474},
  ISSN={1941-0077},
  month={},}

@INPROCEEDINGS{10555071,
  author={Pitsilou, Vasiliki and Papadakis, George and Skoutas, Dimitrios},
  booktitle={2024 IEEE 40th International Conference on Data Engineering Workshops (ICDEW)}, 
  title={Using LLMs to Extract Food Entities from Cooking Recipes}, 
  year={2024},
  volume={},
  number={},
  pages={21-28},
  doi={10.1109/ICDEW61823.2024.00008},
  ISSN={2473-3490},
  month={May},}

@inproceedings{10.1145/3607828.3617797,
author = {Rodr\'{\i}guez-de-Vera, Jes\'{u}s M. and Villacorta, Pablo and Estepa, Imanol G. and Bola\~{n}os, Marc and Saras\'{u}a, Ignacio and Nagarajan, Bhalaji and Radeva, Petia},
title = {Dining on Details: LLM-Guided Expert Networks for Fine-Grained Food Recognition},
year = {2023},
isbn = {9798400702846},
publisher = {Association for Computing Machinery},
address = {New York, NY, USA},
url = {https://doi.org/10.1145/3607828.3617797},
doi = {10.1145/3607828.3617797},
booktitle = {Proceedings of the 8th International Workshop on Multimedia Assisted Dietary Management},
pages = {43–52},
numpages = {10},
location = {Ottawa ON, Canada},
series = {MADiMa '23}
}

@misc{zhou2024foodskyfoodorientedlargelanguage,
      title={FoodSky: A Food-oriented Large Language Model that Passes the Chef and Dietetic Examination}, 
      author={Pengfei Zhou and Weiqing Min and Chaoran Fu and Ying Jin and Mingyu Huang and Xiangyang Li and Shuhuan Mei and Shuqiang Jiang},
      year={2024},
      eprint={2406.10261},
      archivePrefix={arXiv},
      primaryClass={cs.CL},
      url={https://arxiv.org/abs/2406.10261}, 
}

@inproceedings{tan2019efficientnet,
  title={Efficientnet: Rethinking model scaling for convolutional neural networks},
  author={Tan, Mingxing and Le, Quoc},
  booktitle={International conference on machine learning},
  pages={6105--6114},
  year={2019},
  organization={PMLR}
}

@misc{edamam_api,
  author = {{Edamam Inc.}},
  title = {Edamam API},
  year = {2024},
  url = {https://www.edamam.com},
  note = {Accessed: 2024-07-18}
}

@misc{openai_gpt35,
  author = {{OpenAI}},
  title = {GPT-3.5 Turbo Fine-Tuning and API Updates},
  year = {2024},
  url = {https://openai.com/index/gpt-3-5-turbo-fine-tuning-and-api-updates/},
  note = {Accessed: 2024-07-18}
}

@article{krizhevsky2012imagenet,
  title={Imagenet classification with deep convolutional neural networks},
  author={Krizhevsky, Alex and Sutskever, Ilya and Hinton, Geoffrey E},
  journal={Advances in neural information processing systems},
  volume={25},
  year={2012}
}

@article{simonyan2014very,
  title={Very deep convolutional networks for large-scale image recognition},
  author={Simonyan, Karen and Zisserman, Andrew},
  journal={arXiv preprint arXiv:1409.1556},
  year={2014}
}

@inproceedings{he2016deep,
  title={Deep residual learning for image recognition},
  author={He, Kaiming and Zhang, Xiangyu and Ren, Shaoqing and Sun, Jian},
  booktitle={Proceedings of the IEEE conference on computer vision and pattern recognition},
  pages={770--778},
  year={2016}
}

@misc{mongodb,
  author = {{MongoDB, Inc.}},
  title = {MongoDB},
  year = {2024},
  url = {https://www.mongodb.com/},
  note = {Accessed: 2024-07-18}
}

@inproceedings{klasson2019hierarchical,
  title={A Hierarchical Grocery Store Image Dataset with Visual and Semantic Labels},
  author={Klasson, Marcus and Zhang, Cheng and Kjellstr{\"o}m, Hedvig},
  booktitle={IEEE Winter Conference on Applications of Computer Vision (WACV)},
  year={2019}
}

@article{muresan2018fruit,
  title={Fruit recognition from images using deep learning},
  author={Muresan, Horea and Oltean, Mihai},
  journal={Acta Universitatis Sapientiae, Informatica},
  volume={10},
  number={1},
  pages={26--42},
  year={2018}
}

@inproceedings{ming2018food,
  title={Food photo recognition for dietary tracking: System and experiment},
  author={Ming, Zhao-Yan and Chen, Jingjing and Cao, Yu and Forde, Ciar{\'a}n and Ngo, Chong-Wah and Chua, Tat Seng},
  booktitle={MultiMedia Modeling: 24th International Conference, MMM 2018, Bangkok, Thailand, February 5-7, 2018, Proceedings, Part II 24},
  pages={129--141},
  year={2018},
  organization={Springer}
}

@article{gao2019hierarchical,
  title={Hierarchical attention network for visually-aware food recommendation},
  author={Gao, Xiaoyan and Feng, Fuli and He, Xiangnan and Huang, Heyan and Guan, Xinyu and Feng, Chong and Ming, Zhaoyan and Chua, Tat-Seng},
  journal={IEEE Transactions on Multimedia},
  volume={22},
  number={6},
  pages={1647--1659},
  year={2019},
  publisher={IEEE}
}

@article{liu2026profile,
  title={Profile-Aware Trustworthy Recipe Generation with Planner-Critic Agentic Remediation},
  author={Liu, Shanhong and Ng, Pai Chet and Plataniotis, Konstantinos N},
  journal={arXiv preprint arXiv:2610.02969},
  year={2026}
}

\end{document}